\documentclass[twocolumn,showpacs,preprintnumbers,aps]{revtex4}
\usepackage{amsmath}
\usepackage{dcolumn}
\usepackage{bm}
\usepackage{graphicx}
\usepackage{amsfonts}
\usepackage{amssymb}
\usepackage{bbm}
\usepackage{float}
\usepackage[dvipdfm]{hyperref}

\begin{document}

\title{Detecting quantum speedup in closed and open systems}
\author{Zhen-Yu Xu}
\email{zhenyuxu@suda.edu.cn}
\affiliation{College of Physics, Optoelectronics and Energy, Soochow University, Suzhou
215006, China}

\begin{abstract}
We construct a general measure for detecting the quantum speedup in both
closed and open systems. The speed measure is based on the changing rate of
the position of quantum states on a manifold with appropriate monotone
Riemannian metrics. Any increase in speed is a clear signature of dynamical
speedup. To clarify the mechanisms for quantum speedup, we first introduce
the concept of \textit{longitudinal} and \textit{transverse} types of
speedup: the former stems from the time evolution process itself with fixed
initial conditions, while the latter is a result of adjusting initial
conditions. We then apply the proposed measure to several typical closed and
open quantum systems, illustrating that quantum coherence $($or entanglement$%
)$ and the memory effect of the environment together can become resources
for longitudinally or transversely accelerating dynamical evolution under
specific conditions and assumptions.
\end{abstract}

\maketitle

\section{Introduction}

The speed of quantum evolution determines how long it will take for a
quantum state to evolve to a target state in a given process. The shortest
possible time for this dynamical process, characterized by Mandelstam-Tamm
\cite{MT} or Margolus-Levitin lower bound \cite{ML}, is typically called the
quantum speed limit (QSL) time, and this quantity plays an important and
fundamental role in the study of quantum computation \cite{QC2}, quantum
metrology \cite{metrology,metrology1}, quantum thermodynamics \cite%
{thermo,thermo2}, and quantum control \cite{control1,control2}. Recent
decades have witnessed considerable research on the QSL time, both in closed
systems \cite{c1,AU,LV,c2,en1,c3,c3.5,en2,en3,c4,c5,c6,en4,c7,c8,c8.5,c9}
and in more general open systems \cite%
{s1,s2,QSLopen1,QSLopen2,QSLopen3,QSLexp,QSLopen4,k7,k1,Xu1,Zhang,Xu2,Wu,Sun,JJ,Xu4,Wei,KP}%
. One important discovery is the fact that entanglement is able to speed up
the evolution of closed quantum systems \cite{en1,en2,en3,en4}. Further
studies have shown that the memory effect of the environment can also induce
dynamical acceleration in open quantum systems \cite{QSLopen3,Xu1}; this
phenomenon has been experimentally observed in recent studies with a cavity
field as the open system and the number of atoms as the controllable
environment \cite{QSLexp}. The above discoveries are of considerable
interest and importance because they reveal that both entanglement and the
memory effect may be beneficial for controlling dynamical processes.

However, as noted in Ref. \cite{Xu2}, no consistent definition exists for
dynamical speedup, which has been conceptualized from multiple different
perspectives. For instance, in the study of the entanglement-assisted
speedup of the quantum evolution of closed systems, the primary focus is on
the actual evolution time $\tau $. Speedup occurs when the actual driving
time $\tau $ approaches the QSL time $\tau _{QSL}$, i.e., $\tau /\tau
_{QSL}\rightarrow 1$ \cite{en1,en2,en3,en4}$.$ From another point of view,
we may focus on the QSL time itself, with the actual driving time $\tau $
held fixed. In this case, $\tau /\tau _{QSL}=1$ implies that the evolution
has already been proceeding along the fastest possible path and that there
is no capacity for further speedup, whereas for $\tau /\tau _{QSL}>1,$ the
system is thought to possess greater capacity for potential speedup \cite%
{QSLopen3,QSLexp,Xu1,Zhang,Xu2,Wu,Sun,Xu4,Wei}. Therefore, there is a strong
need for a well-defined measure of the speed of quantum evolution that can
be employed both to detect real (not just potential) dynamical speedup and
to reexamine the mechanisms of dynamical speedup in closed and open systems.

Here, we construct such a measure for detecting quantum speedup. Analogous
to the instantaneous speed of an object based on the rate of change of its
position in Euclidean space as calculated using the Euclidean metric, the
speed of quantum evolution is defined as the changing rate of the position
of quantum states on a manifold with appropriate monotone Riemannian
metrics. The quantum speedup is then detected by any increase of the speed.
To further clarify the mechanisms for the quantum speedup, the
\textquotedblleft speedup\textquotedblright\ is divided into two types,
i.e., \textit{longitudinal\quad speedup }and\textit{\ transverse speedup}.
The former arises during the evolution process itself with fixed initial
conditions, while the latter is due to a change in the initial conditions.
We then apply the proposed measure to several typical closed and open
quantum systems to illustrate the phenomena of longitudinal and transverse
speedup. We find that quantum coherence and entanglement can serve as
resources for such speedup only for specific types of quantum states in
closed and open systems. In addition, the memory effect in the given
examples is also confirmed to be a subtle but important factor in
determining the longitudinal or transverse acceleration of the evolution
process.

The paper is organized as follows: In Sec. II, we construct the measure for
detecting quantum speedup based on appropriate monotone Riemannian metrics.
The mechanism for longitudinal and transverse quantum speedup in closed and
open systems is analyzed in Sec. III with two typical and pedagogical
examples. Finally, conclusions are drawn in Sec. IV.

\section{Detection of the speedup of dynamical evolution}

To construct the measure of dynamical speed, let us first recall several
notations for quantum states. The physical state of a quantum system is
represented by the density operator $\rho $, and the set of density
operators $M_{n}$ that we consider here is a subset of bounded operators $B(%
\mathcal{H})$ defined on the finite $n$-dimensional Hilbert space $\mathcal{H%
}$. Given the initial state $\rho _{0}$, the dynamics is governed by $\rho
_{t}=\Lambda _{t}\rho _{0}$, where $\Lambda _{t}$ is the dynamical map \cite%
{Book Open}. From the perspective of the differential geometry of quantum
states, the evolved density operators constitute a manifold $\mathcal{M}_{n}$%
, and the dynamical map $\Lambda _{t}$ can be treated as a curve on $%
\mathcal{M}_{n}$ with parameter $t$ \cite{book-gs}. We note that at every
time point $t$, the quantum state possesses its own tangent vector space $T$
on $M_{n}$, which constitutes a tangent vector field $\mathcal{F}_{T}$ on
the manifold $\mathcal{M}_{n}$ when the entire process of quantum evolution
is considered. To obtain the length of the curve, an appropriate monotone
Riemannian metric tensor field $g:$ $\mathcal{F}_{T}\times \mathcal{F}%
_{T}\rightarrow F$ (where $F$ denotes the scalar field on $\mathcal{M}_{n}$)
should be given. Then, the line element of the curve is defined as [Note 1]
\begin{equation}
ds^{2}:=(dl)^{2}\text{ with }dl:=\left\Vert \dot{\rho}_{t}\right\Vert dt,
\label{line element}
\end{equation}%
where $\left\Vert X\right\Vert :=\sqrt{g(X,X)}$ denotes the length of any
tangent vector $X\in T$ and $\dot{\rho}_{t}:=\partial _{t}\rho _{t}\in
\mathcal{F}_{T}$ \cite{book-gs,book-dg}$.$ The length of the curve is then
expressed as
\begin{equation}
l:=\int dl=\int_{0}^{t}\sqrt{g\left( \partial _{_{t^{\prime }}}\rho
_{t^{\prime }},\partial _{_{t^{\prime }}}\rho _{t^{\prime }}\right) }%
dt^{^{\prime }}
\end{equation}%
The above construction allows us to define the instantaneous speed of
quantum evolution as

\begin{equation}
S:=\dot{l}=\sqrt{g\left( \dot{\rho}_{t},\dot{\rho}_{t}\right) }.
\end{equation}

To derive an explicit form of the above quantity, we must select an
appropriate monotone Riemannian metric. If the quantum state is pure, then
the Fubini-Study metric is the natural choice. However, if the quantum
system exists in mixed states, then there are infinitely many monotone
Riemannian metrics \cite{book-gs}. According to the theorem of Morozova,
Chentsov, and Petz, any monotone Riemannian metric can be represented in the
unified form $g(X,Y):=\frac{1}{4}\left\langle X,K^{-1}(Y)\right\rangle $,
with $X,Y\in T$. Here $\left\langle \cdot ,\cdot \right\rangle $ denotes the
Hilbert-Schmidt inner product, and $K$ is a positive superoperator defined
as $K:=f(L_{\rho }R_{\rho }^{-1})R_{\rho }$, where $R_{\rho }(X):=X\rho $
and $L_{\rho }(X):=\rho X$, and $f:%
\mathbb{R}
^{+}\rightarrow
\mathbb{R}
^{+}$ is an operator monotone function $f(x)=xf(x^{-1})$ being normalized by
$f(1)=1$ \cite{petz}. Therefore, the speed of quantum evolution is given by%
\begin{equation}
S=\frac{1}{2}\sqrt{\left\langle \dot{\rho}_{t},K^{-1}(\dot{\rho}%
_{t})\right\rangle }.  \label{speed-0}
\end{equation}%
To obtain the above measure in an explicit form, we may rewrite the density
operator in the form of its spectral decomposition $\rho
_{t}=\sum_{k}p_{k}\left\vert \Phi _{k}\right\rangle \left\langle \Phi
_{k}\right\vert $, with $0<p_{k}<1$ and $\sum_{k}p_{k}=1.$ In the
Morozova-Chentsov-Petz formalism \cite{book-gs}, Eq. (\ref{speed-0}) can be
immediately written as%
\begin{equation}
S=\frac{1}{2}\sqrt{\sum_{k,l}c(p_{k},p_{l})\left\vert \langle \Phi _{k}|\dot{%
\rho}_{t}|\Phi _{l}\rangle \right\vert ^{2}}\text{,}  \label{Speed-1}
\end{equation}%
where the symmetric function $c(x,y)=1/[yf(x/y)]$ is the so-called
Morozova-Chentsov function \cite{petz} related to our chosen Riemannian
metrics.

We note that any monotone Riemannian metric can be employed to evaluate the
instantaneous speed of quantum evolution with Eq. (\ref{Speed-1}) as long as
the quantum state is in the interior of the space of quantum states, i.e., $%
0<p_{k}<1.$ However, because a measure of speed should be applicable to both
closed and open systems, it is hoped that Eq. (\ref{Speed-1}) could be
continuously extended to the boundary of the manifold, i.e., one or more of $%
p_{k}$ vanishes. As can be seen from Eq. (\ref{Speed-1}), two kinds of
singularity, i.e., $c(p_{k},p_{k})=1/p_{k}\rightarrow \infty $ and $%
c(p_{k},p_{l})\rightarrow \infty $ ($k\neq l)$, may appear when we are
trying to make an extension to the boundary. Fortunately, the first
singularity can be removed by employing a strategy of replacing $p_{k}$\
with $q_{k}$ ($q_{k}^{2}=p_{k})$ \cite{book-gs}. However, to our knowledge,
there exists only a few monotone Riemannian metrics being able to avoid the
second kind of singularity [Note2], such as the symmetric logarithmic
derivative metric, with $c_{SL}(x,y)=\frac{2}{x+y}$, and the Wigner-Yanase
metric, with $c_{WY}(x,y)=\frac{4}{(\sqrt{x}+\sqrt{y})^{2}}$. The above
consideration is justified when we take the pure state $\rho _{t}=\left\vert
\psi \right\rangle \left\langle \psi \right\vert $ as an example, whose line
element is the well known $ds_{FS}^{2}=\langle d\psi _{\perp }|d\psi _{\perp
}\rangle $ of the Fubini-Study metric \cite{s1,book-gs}. Here $\left\vert
d\psi _{\perp }\right\rangle :=\left\vert d\psi \right\rangle -\langle \psi
|d\psi \rangle \left\vert \psi \right\rangle $ is the component of $%
\left\vert d\psi \right\rangle $ orthogonal to $\left\vert \psi
\right\rangle $. It is convenient to check that, for pure states, the line
elements of the symmetric logarithmic derivative metric and the
Wigner-Yanase metric will reduce to $ds_{SL}^{2}=\langle d\psi _{\perp
}|d\psi _{\perp }\rangle =ds_{FS}^{2}$ and $ds_{WY}^{2}=2\langle d\psi
_{\perp }|d\psi _{\perp }\rangle =2ds_{FS}^{2}$, respectively.

In terms of above strategy and selected metrics, we have
\begin{equation}
S=\sqrt{\sum_{k}\dot{q}_{k}^{2}+\sum_{k\neq l}c_{SL(WY)}(p_{k},p_{l})\frac{%
p_{k}(p_{k}-p_{l})}{2}\left\vert \left\langle \Phi _{l}\right\vert \dot{\Phi}%
_{k}\rangle \right\vert ^{2}},  \label{speed-A}
\end{equation}%
where we have utilized $c(x,y)=c(y,x)$ and $\left\langle \Phi
_{l}\right\vert \dot{\Phi}_{k}\rangle +\langle \dot{\Phi}_{l}|\Phi
_{k}\rangle =0.$ As a special case when the system is a closed system with a
pure state $\rho _{t}=\left\vert \psi \right\rangle \left\langle \psi
\right\vert $, it is simple to check that Eq. (\ref{speed-A}) reduces to%
\begin{equation}
S=\frac{\epsilon \left\vert \langle d\psi _{\perp }|\dot{\psi}\rangle
\right\vert }{\sqrt{\langle d\psi _{\perp }|d\psi _{\perp }\rangle }},
\label{speed-B}
\end{equation}%
with $\epsilon =1,\sqrt{2}$ for the symmetric logarithmic derivative metric
and the Wigner-Yanase metric, respectively. In the following, we will focus
on the symmetric logarithmic derivative metric as an example; extension to
the Wigner-Yanase metric and other potential appropriate metrics \cite%
{book-distance} is straightforward.

Equation (\ref{speed-A}) can now serve as a basis for detecting quantum
speedup. Clearly, any increase in speed is a signature of dynamical speedup:
\begin{equation}
\partial _{\xi }S>0,  \label{speed up}
\end{equation}%
where $\xi $ is the concerned dynamical parameter. In the following, Eq. (%
\ref{speed up}) (or $\partial _{\xi }S$) will be employed as a detection (or
a measure) for quantum speedup. To clarify the mechanisms underlying speedup
phenomena, it is of necessity to distinguish two types of \textquotedblleft
speedup\textquotedblright : (i) \textit{longitudinal\quad speedup }(i.e., $%
\xi =t$), which arises during the evolution process itself with certain
initial states and fixed environmental parameters, and (ii) \textit{%
transverse speedup} (i.e., $\xi \in \{\xi _{1},\xi _{2},\xi _{3},\cdots \}$
and $\partial _{t}\xi _{i}=0$) due to a change in the initial conditions
(parameterized by $\xi _{i}$), such as the initial states and environmental
parameters.

\section{Examples}

In what follows, we apply the measure constructed above to several simple
but pedagogical examples and illustrate the mechanisms corresponding to both
longitudinal speedup and transverse speedup.

\subsection{Closed systems: a physical model of spin precession}

\subsubsection{Single-qubit case}

Consider a physical model of spin precession, with a spin-$\frac{1}{2}$
system subjected to a uniform static external magnetic field in the $z$
direction. The Hamiltonian of the system can be written as \cite%
{book-quantum} ($\hbar =1$)
\begin{equation}
H=\frac{\omega }{2}\sigma _{z},  \label{H1}
\end{equation}%
where $\omega $ is the energy difference between the two spin eigenstates
and $\sigma _{z}:=\left\vert 1\right\rangle \left\langle 1\right\vert
-\left\vert 0\right\rangle \left\langle 0\right\vert $ is the usual Pauli
operator. If the system is initially prepared in the superposition state $%
\left\vert \psi _{0}\right\rangle =\alpha \left\vert 1\right\rangle +\beta
\left\vert 0\right\rangle $, with $|\alpha |^{2}+|\beta |^{2}=1,$ then the
time evolution of the spin is given by $\left\vert \psi _{t}\right\rangle
=\alpha \exp (-i\omega t/2)\left\vert 1\right\rangle +\beta \exp (i\omega
t/2)\left\vert 0\right\rangle $. Using Eq. (\ref{speed-B}), we have
\begin{equation}
S=\left\vert \alpha \beta \right\vert \omega .  \label{S1}
\end{equation}%
Therefore, transverse speedup (i.e., $\partial _{\alpha }S>0$) can be
achieved by enhancing the initial coherence of the spin state. With $\alpha $%
($\beta $) given initially in Equation (\ref{S1}), the system will evolve at
a uniform speed (i.e., longitudinal speedup will never occur, for $\partial
_{t}S\equiv 0$). In particular, the speed of evolution will be zero when $%
\alpha (\beta )=0$. It is easily understood, for example, that if the system
is initially in state $\left\vert 1\right\rangle $, then the evolved state
will be $\exp (-i\omega t/2)\left\vert 1\right\rangle $, which is precisely
identical to $\left\vert 1\right\rangle $. In fact, both state $\left\vert
1\right\rangle $ and state $\exp (-i\omega t/2)\left\vert 1\right\rangle $
belong to the same ray in projective Hilbert space; hence, the system has
not evolved at all.

\subsubsection{Two-qubit case}

The previous single-qubit case can be generalized to a bipartite system.
Under the assumption that no interaction exists between the two spins, the
evolution is governed by the following Hamiltonian:

\begin{equation}
H=H^{A}\otimes I^{B}+I^{A}\otimes H^{B}  \label{H2}
\end{equation}%
with $H^{A(B)}=$ $\omega \sigma _{z}^{A(B)}/2$, where $I^{A(B)}$ is the
identity operator of the spin $A(B)$. For convenience and without loss of
generality, the initial state is first set to $\left\vert \varphi
_{0}\right\rangle =\alpha \left\vert 11\right\rangle +\beta \left\vert
00\right\rangle $. The corresponding evolved state is then written as $%
\left\vert \varphi _{t}\right\rangle =\alpha \exp (-i\omega t)\left\vert
11\right\rangle +\beta \exp (i\omega t)\left\vert 00\right\rangle $.
According to Eq. (\ref{speed-B}), we obtain

\begin{equation}
S=2\left\vert \alpha \beta \right\vert \omega ,  \label{S2-1}
\end{equation}%
which implies that the longitudinal speedup (i.e., $\partial _{t}S\equiv 0$)
will never occur, but the transverse speedup (i.e., $\partial _{\alpha }S>0$%
) may be realized by adjusting the initial states. In fact, the speed in
equation (\ref{S2-1}) is related to the entanglement of the initially
prepared state. To demonstrate this fact, we employ Wootter's concurrence $C$
to measure the entanglement of the bipartite system \cite{concurrence}.
Because the evolved density operator is of the \textquotedblleft X" type,
the concurrence can be easily deduced to be $C=2|\alpha \beta |$ (related to
initial state). Hence, the evolution speed is given by%
\begin{equation}
S=C\omega ,  \label{S2-2}
\end{equation}%
illustrating an interesting phenomenon in which quantum entanglement is able
to induce transverse speedup (i.e., $\partial _{C}S=\omega >0$) through
control of the initial states.

Nevertheless, we note that not all entangled states possess such a property,
e.g., if the system is initially prepared in the state $\left\vert \phi
_{0}\right\rangle =\alpha \left\vert 10\right\rangle +\beta \left\vert
01\right\rangle $, it is easy to check, using Eq. (\ref{speed-B}), that $S=0$
because $\left\vert \phi _{t}\right\rangle \equiv \left\vert \phi
_{0}\right\rangle $. However, the entanglement of this state is always $%
C=2\left\vert \alpha \beta \right\vert $ and is fully uncorrelated with the
speed of quantum evolution. Therefore, the entanglement induced transverse
speedup is fully state-dependent in this example.

\subsection{Open systems: a physical model of two-level atoms coupled to
leaky cavities}

\subsubsection{Single-qubit case}

\begin{figure}[]
\centering{}\includegraphics[width=3.5in]{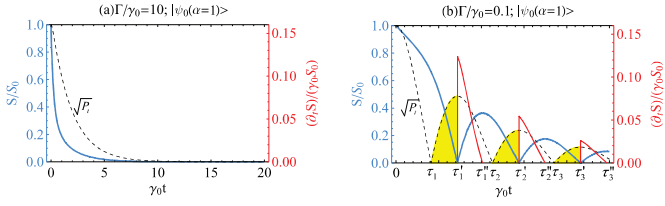}
\caption{Comparison among the normalized speed of quantum evolution, $%
S/S_{0} $ (blue curve); the detection for memory effects, $\protect\sqrt{%
P_{t}}$ (dashed curve); and the measure for longitudinal speedup, $(\partial
_{t}S)/(\protect\gamma _{0}S_{0})$ (red curve), as functions of $\protect%
\gamma _{0}t $, for a singe-qubit open system with fixed initial state $%
\left\vert \protect\psi _{0}(\protect\alpha =1)\right\rangle =\left\vert
1\right\rangle $ in (a) a memoryless environment $(\Gamma /\protect\gamma %
_{0}=10)$, and (b) a memory environment $(\Gamma /\protect\gamma _{0}=0.1)$,
respectively. From a phenomenological point of view, the longitudinal
speedup in (b) within $\protect\gamma _{0}t\in \lbrack \protect\tau %
_{n}^{\prime },\protect\tau _{n}^{\prime \prime }]$ is driven by the
previous accumulation of memory effects (yellow-shaded regions within $%
\protect\gamma _{0}t\in \lbrack \protect\tau _{n},\protect\tau _{n}^{\prime
}]$).}
\end{figure}

We now turn to the case of an open system. Consider a two-level atom coupled
to a leaky vacuum cavity with the following Hamiltonian \cite{Book Open}:

\begin{equation}
H=\omega _{0}\sigma _{+}\sigma _{-}+\sum_{k}\omega _{k}a_{k}a_{k}^{\dag
}+i\sum_{k}\zeta _{k}(a_{k}^{\dag }\sigma _{-}-a_{k}\sigma _{+}),  \label{H3}
\end{equation}%
where $\omega _{0}$ is the resonant transition frequency of the atom between
the ground state $|0\rangle $ and the excited state $|1\rangle $; $\sigma
_{+}:=\left\vert 1\right\rangle \left\langle 0\right\vert $ and $\sigma
_{-}:=\left\vert 0\right\rangle \left\langle 1\right\vert $ are the Pauli
raising and lowering operators, respectively; and $\omega _{k}$ and $a_{k}$ (%
$a_{k}^{\dag }$) denote the frequency and annihilation (creation) operator,
respectively, of the $k$th cavity mode, with $\zeta _{k}$ being the
corresponding coupling constant. The reduced density matrix of the atom,
with an initial state $\rho =(\rho _{mn})$ ($m,n=0,1$), takes the form \cite%
{Book Open}

\begin{equation}
\rho _{t}=\left(
\begin{array}{cc}
\rho _{11}P_{t} & \rho _{10}\sqrt{P_{t}} \\
\rho _{01}\sqrt{P_{t}} & 1-\rho _{11}P_{t}%
\end{array}%
\right) ,  \label{exp3-rou}
\end{equation}%
where $P_{t}$ is related to the excited state population of the atom. For
arbitrary initial states, the speed of quantum evolution can be acquired by
Eq. (\ref{speed-A}), but in most cases, only numerical calculation is
feasible. For simplicity, here, we consider the speed of evolution with an
initial state of $\left\vert \psi _{0}\right\rangle =\alpha \left\vert
1\right\rangle +\sqrt{1-\alpha ^{2}}\left\vert 0\right\rangle $ ($\alpha \in
\lbrack 0,1]$). Then, the spectral decomposition of Eq. (\ref{exp3-rou}) can
be obtained by $\rho _{t}=\sum_{k=\mp }p_{k}\left\vert \Phi
_{k}\right\rangle \left\langle \Phi _{k}\right\vert $ with $p_{\mp }=(1\mp
\eta )/2$; $\left\vert \Phi _{\mp }\right\rangle =(b_{\mp }\left\vert
1\right\rangle +\left\vert 0\right\rangle )/\sqrt{1+b_{\mp }^{2}},$ where $%
\eta =\sqrt{1-4\alpha ^{4}P_{t}+4\alpha ^{4}P_{t}^{2}}$ and $b_{\mp
}=-(1-2\alpha ^{2}P_{t}\pm \eta )/(2\alpha \sqrt{1-\alpha ^{2}}\sqrt{P_{t}}%
). $ According to Eq. (\ref{speed-A}), we have

\begin{equation}
S=\frac{\alpha \left\vert \dot{P}_{t}\right\vert }{2}\sqrt{\frac{1-(1-\alpha
^{2})P_{t}}{P_{t}(1-P_{t})}}.  \label{S3}
\end{equation}

To be specific, let us examine a switchable environment with the Lorentzian
spectral density $J(\omega )=\frac{1}{2\pi }\frac{\gamma _{0}\Gamma }{%
(\omega _{0}-\omega )^{2}+\Gamma ^{2}}$, where $\gamma _{0}$ is the
Markovian-limit decay rate and $\Gamma $ is the spectral width \cite{Book
Open}. Hence, $P_{t}=e^{-\Gamma t}[\cos (\frac{\kappa t}{2})+\frac{\Gamma }{%
\kappa }\sin (\frac{\kappa t}{2})]^{2}$ with $\kappa =\sqrt{2\gamma
_{0}\Gamma -\Gamma ^{2}}$, which can be controlled by the width of the
spectrum: $\Gamma /\gamma _{0}>2$ represents the memoryless (Markovian)
region, and $\Gamma /\gamma _{0}<2$ corresponds to the non-Markovian region
with memory. The maximal speed occurs at $t=0$, with
\begin{equation}
S_{0}:=\lim_{t\rightarrow 0}S=\alpha ^{2}\sqrt{\frac{\Gamma \gamma _{0}}{2}}.
\label{max speed}
\end{equation}

Obviously, it can be confirmed that if $\Gamma /\gamma _{0}>2$, then Eq. (%
\ref{S3}) decreases monotonically to zero, implying that no longitudinal
speedup occurs during the evolution [e.g., see in Fig. 1(a)]. However, if $%
\Gamma /\gamma _{0}<2$, then longitudinal speedup will occur. To illustrate
this phenomenon, the normalized speed of quantum evolution, $S/S_{0}$ (blue
curve); and the measure for quantum speedup, ($\partial _{t}S$)/($\gamma
_{0}S_{0}$) (red curve) with $\Gamma /\gamma _{0}=0.1$ and $\alpha =1$
(i.e., $\left\vert \psi _{0}(\alpha =1)\right\rangle =\left\vert
1\right\rangle $) is depicted in Fig. 1(b). The longitudinal speedup ($%
\partial _{t}S>0$) regions are within $\gamma _{0}t\in \lbrack \tau
_{n}^{\prime },\tau _{n}^{\prime \prime }]$ ($n\in
\mathbb{Z}
^{+}$), where $\tau _{n}^{\prime }/\gamma _{0}=2n\pi /\kappa $ and $\tau
_{n}^{\prime \prime }/\gamma _{0}$ denote the solutions to the
transcendental equation $\Gamma \tan (\kappa t/2)=\kappa \tanh (\Gamma t/2)$%
. This phenomenon can be phenomenologically explained in terms of the memory
effects of the environment \cite{nonM-review1,nonM-review2}. The memory
effect of the above model is detected by $\sqrt{P_{t}}$ \cite{Xu3}, which is
drawn as the dashed curve in Fig. 1(b), with the memory regions (marked in
shades of yellow) corresponding to $\gamma _{0}t\in \lbrack \tau _{n},\tau
_{n}^{\prime }]$ ($n\in
\mathbb{Z}
^{+})$, where $\tau _{n}/\gamma _{0}=2$\textbf{[}$n\pi -\arctan (\kappa
/\Gamma )$\textbf{]}$/\kappa $. From a phenomenological perspective, the
longitudinal speedup is driven by the accumulation of the previous memory
effect of the environment.

\begin{figure}[]
\centering{}\includegraphics[width=3.5in]{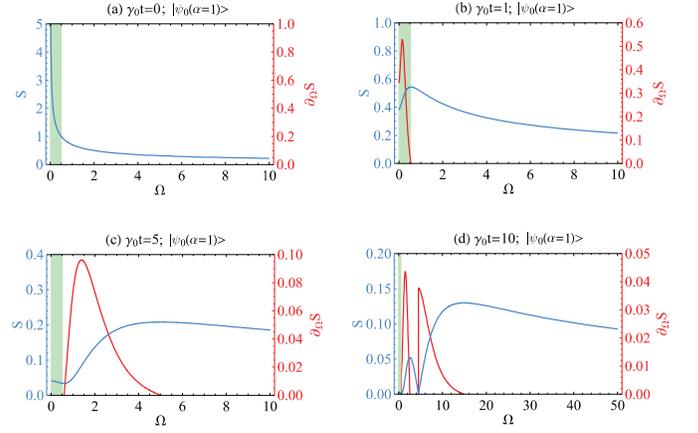}
\caption{Comparison between the speed of quantum evolution, $S$ (blue
curve); and the measure for transverse speedup, $\partial _{\Omega }S$ (red
curve), as functions of $\Omega $ (=$\protect\gamma _{0}/\Gamma $), for a
single-qubit open system prepared in initial state $\left\vert \protect\psi %
_{0}(\protect\alpha =1)\right\rangle =\left\vert 1\right\rangle $ with fixed
point in time at (a) $\protect\gamma _{0}t=0$; (b) $\protect\gamma _{0}t=1$;
(c) $\protect\gamma _{0}t=5$; and (d) $\protect\gamma _{0}t=10$,
respectively. A narrower spectral width $\Gamma $ (i.e., larger $\Omega $)
corresponds to a stronger memory effect [$\Omega \in (0,1/2)$ for Markovian
(green-shaded) region, and $\Omega \in (1/2,\infty )$ for non-Markovian
region].}
\end{figure}

Meanwhile, to study the memory effect on the transverse speedup, we can
select the driving time and then adjust the initial parameters of the
environment. To be clear, the speed of quantum evolution, $S$ (blue curve);
and the measure for transverse speedup, $\partial _{\Omega }S$ (red curve)
are plotted in Fig. 2 versus $\Omega $ (=$\gamma _{0}/\Gamma $), with fixed
driving time. As clearly shown in Fig. 2(b), transverse speedup ($\partial
_{\Omega }S>0$) may occur even in the Markovian region [green shaded, $%
\Omega \in (0,1/2)$]. When the driving time is longer [e.g., Fig. 2(c) and
(d)], the transverse speedup ($\partial _{\Omega }S>0$) will take place in
the non-Markovian region [$\Omega \in (1/2,\infty )$].

\subsubsection{Two-qubit case}

Consider the case of a bipartite system of independent two-level atoms, each
locally coupled to a leaky vacuum cavity. The dynamics of this open system
is determined by each atom-cavity pair in the same manner illustrated in the
single-qubit case. Consider, for instance and simplicity, the entangled
initial state $\left\vert \varphi _{0}\right\rangle =\alpha \left\vert
11\right\rangle +\sqrt{1-\alpha ^{2}}\left\vert 00\right\rangle $ ($\alpha
\in \lbrack 0,1]$); the evolved state can be immediately obtained \cite%
{Bellomo}:
\begin{equation}
{\tiny {{\rho _{t}=\left(
\begin{array}{llll}
\alpha ^{2}P_{t}^{2} & 0 & 0 & \alpha \sqrt{1-\alpha ^{2}}P_{t} \\
0 & \alpha ^{2}P_{t}(1-P_{t}) & 0 & 0 \\
0 & 0 & \alpha ^{2}P_{t}(1-P_{t}) & 0 \\
\alpha \sqrt{1-\alpha ^{2}}P_{t} & 0 & 0 & 1-2\alpha ^{2}P_{t}+\alpha
^{2}P_{t}^{2}%
\end{array}%
\right) .}}}  \label{exp4-rou}
\end{equation}%
Then, the spectral decomposition of Eq. (\ref{exp4-rou}) is obtained by $%
\rho _{t}=\sum_{k=1,2,\mp }p_{k}\left\vert \Phi _{k}\right\rangle
\left\langle \Phi _{k}\right\vert $ with $p_{1}=p_{2}=\alpha
^{2}P_{t}(1-P_{t}),$ $p_{\mp }=(1-2\alpha ^{2}P_{t}+2\alpha ^{2}P_{t}^{2}\mp
\delta )/2;$ $\left\vert \Phi _{1}\right\rangle =\left\vert 01\right\rangle
, $ $\left\vert \Phi _{2}\right\rangle =\left\vert 10\right\rangle ,$ $%
\left\vert \Phi _{\mp }\right\rangle =(d_{\mp }\left\vert 11\right\rangle
+\left\vert 00\right\rangle )/\sqrt{1+d_{\mp }^{2}}$, where $\delta =\sqrt{%
1-4\alpha ^{2}P_{t}+4\alpha ^{2}P_{t}^{2}},$ $d_{\mp }=-(1-2\alpha
^{2}P_{t}\pm \delta )/(2\alpha \sqrt{1-\alpha ^{2}}P_{t}).$ According to Eq.
(\ref{speed-A}), we have
\begin{equation}
S=\alpha |\dot{P}_{t}|\sqrt{\frac{1-2P_{t}+2P_{t}^{2}}{2P_{t}(1-P_{t})(1-2%
\alpha ^{2}P_{t}+2\alpha ^{2}P_{t}^{2})}}.  \label{S4}
\end{equation}

\begin{figure}[]
\centering{}\includegraphics[width=3.5in]{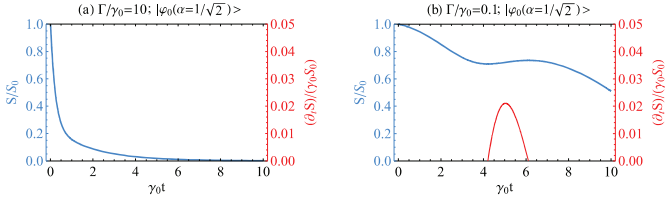}
\caption{{}Comparison between the normalized speed of quantum evolution, $%
S/S_{0}$ (blue curve); and the measure for longitudinal speedup, $(\partial
_{t}S)/(\protect\gamma _{0}S_{0})$ (red curve), as functions of $\protect%
\gamma _{0}t$, for a two-qubit open system initially prepared in state $%
\left\vert \protect\varphi _{0}(\protect\alpha =1/\protect\sqrt{2}%
)\right\rangle =(\left\vert 11\right\rangle +\left\vert 00\right\rangle )/%
\protect\sqrt{2}$; and surrounded by (a) a memoryless environment $(\Gamma /%
\protect\gamma _{0}=10)$, and (b) a memory environment $(\Gamma /\protect%
\gamma _{0}=0.1)$, respectively. }
\end{figure}
It can be confirmed that the maximal instantaneous speed also occurs at $t=0$
with $S_{0}=\alpha \sqrt{\Gamma \gamma _{0}}$. In addition, Eq. (\ref{S4})
also implies that longitudinal speedup ($\partial _{t}S>0$) occurs in a
memory environment, precisely as shown in Fig. 3. \

\begin{figure}[]
\centering{}\includegraphics[width=3.5in]{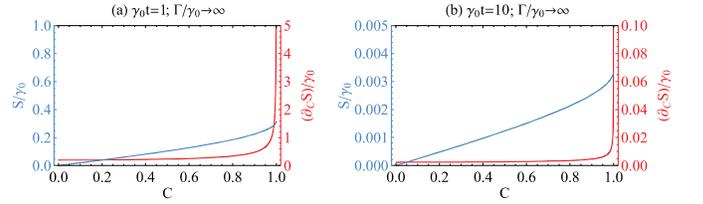}
\caption{Comparison between the speed of quantum evolution, $S/\protect%
\gamma _{0}$ (blue curve); and the measure for transverse speedup, $%
(\partial _{C}S)/\protect\gamma _{0}$ (red curve), as functions of $C$
(entanglement of initial two-qubit state measured by Concurrence) in a
Markovian-limit environment $(\Gamma /\protect\gamma _{0}\rightarrow \infty
) $ with fixed point in time at (a) $\protect\gamma _{0}t=1;$ (b) $\protect%
\gamma _{0}t=10$, respectively.}
\end{figure}

On the other hand, to study the transverse speedup, e.g., the influence of
initial entanglement $C$ ($=2\alpha \sqrt{1-\alpha ^{2}}$) on the speedup,
we may select an environment (e.g., the Markovian limit with $\Gamma
\rightarrow \infty $ and $P_{t}=\exp (-\gamma _{0}t)$], and fix the driving
time $t$. Equation (\ref{S4}) thereby reduces to%
\begin{equation}
\frac{S}{\gamma _{0}}=\frac{1}{2}\sqrt{\frac{(1-\sqrt{1-C^{2}}%
)P_{t}(1-2P_{t}+2P_{t}^{2})}{(1-P_{t})[1-(1-\sqrt{1-C^{2}})P_{t}(1-P_{t})]}}.
\end{equation}%
As is clearly shown in Fig. 4, stronger entanglement of initial state will
give rise to the transverse speedup ($\partial _{C}S>0$).

However, we note that if the system is prepared in the initial state, e.g., $%
\left\vert \phi _{0}\right\rangle =\alpha \left\vert 10\right\rangle +\beta
\left\vert 01\right\rangle $, then we have $S=\frac{|\dot{P}_{t}|}{2\sqrt{%
P_{t}(1-P_{t})}}$ [irrelevant to $\alpha (\beta )$], implying that the
entanglement of initial state is fully uncorrelated with the speed in this
example, i.e., the transverse speedup ($\partial _{C}S>0$) never occurs
under such circumstances, which bears a resemblance to the case of example
A-2.

\section{Conclusion}

We have constructed a well-defined geometrical measure for detecting the
real speedup of quantum evolution. Furthermore, the quantum speedup has been
subdivided into longitudinal and transverse types. The former focuses on the
time evolution process with fixed initial conditions, while the latter
concerns influence of the initial conditions on quantum speedup. With such
constructed measure, the mechanisms for quantum speedup have been explored
within several typical closed and open systems, thereby demonstrating the
fact that quantum coherence/entanglement as well as the memory effect may
serve as resources for longitudinally or transversely accelerating the
evolution of quantum states under specific conditions in these examples.

We have analyzed the mechanisms for both longitudinal and transverse speedup
with simple examples, and it will be of great interest and significance to
perform further investigations on much complex physical systems, such as
controllable atoms \cite{QSLexp}, quantum dots \cite{exp-dot}, or
nitrogen-vacancy centers \cite{exp-nv}\ confined within microcavities.

\section*{NOTES}

Note 1: The notation $ds^{2}$ in differential geometry can be understood
from two different perspectives \cite{book-dg}. In the first, it is
considered to be the metric tensor $g$ itself expressed in the cotangent
vector space. The second relates $ds^{2}$ to square of the length of the
tangent vector. In this work, we adopt the latter view.

Note 2: Other well known metrics such as the right logarithmic derivative
metric and the Bogoliubov-Kubo-Mori metric \cite{book-distance} are not
appropriate under the framework of our proposed measure, for they can not be
continuously extended to the boundary of the manifold.

\section*{ACKNOWLEDGMENTS}

This work was supported by the National Natural Science Foundation of China
(Grant No. 11204196) and the Specialized Research Fund for the Doctoral
Program of Higher Education (Grant No. 20123201120004).

\end{document}